\documentclass[aps,prb,twocolumn,preprintnumbers,amsmath,superscriptaddress,amssymb]{revtex4}

\usepackage{amssymb}
\usepackage{graphicx}
\usepackage{amsmath}
\usepackage{amsfonts}
\usepackage{epsfig}
\usepackage{bm}
\usepackage{color}

\begin{document}

\title{Critical role of the sign
structure in the doped Mott insulator: Superconductivity vs. Fermi liquid}
\author{Hong-Chen Jiang}
\email{hcjiang@stanford.edu}
\affiliation{Stanford Institute for Materials and Energy Sciences, SLAC and Stanford
University, Menlo Park, California 94025, USA}
\author{Shuai Chen}
\affiliation{Institute for Advanced Study, Tsinghua University, Beijing 100084, China}
\author{Zheng-Yu Weng}
\affiliation{Institute for Advanced Study, Tsinghua University, Beijing 100084, China}
\date{\today }

\begin{abstract}
Mechanism of superconductivity (SC) in a purely interacting electron system
has been one of the most challenging issues in condensed matter physics. In
the BCS theory, the Landau's Fermi liquid is a normal state against which an
SC instability occurs once an \emph{additional} pairing force is added. We
show that in the doped Mott insulator an \emph{intrinsic} SC ground state
(specifically a Luther-Emery state in a finite-doping two-leg $t$-$J$
ladder) can be directly turned into a Fermi-gas-like state by merely switching
off the hidden statistical sign structure via two schemes. It points to a
new pairing paradigm, 
which is an ``Amperean-like pairing'' with a ``stringlike'' pairing force as
shown by an adiabatic continuity to a strong anisotropic limit of the model.

\end{abstract}

\maketitle

\emph{Introduction.}---Recently large-scale density matrix renormalization
group (DMRG) calculations have shown that the $t$-$J$ and Hubbard models
on two- and four-leg ladders \cite{Noack1994,Poilblanc1995,Noack1996,Dolfi2015,Jiang2018tj,Jiang2018hub,Jiang2019hub} have a Luther-Emery (LE) ground state \cite{Luther1974} at finite doping, which is a superconducting
(SC) state for the quasi one-dimensional (1D) system. Due to the finite
spin-spin correlation at half-filling, the doped case of an even-leg spin
ladder may provide a prototypical model for examining the underlying SC
mechanism in a doped Mott insulator, especially that of the
resonating-valence-bond (RVB) mechanism\cite{Anderson_1987} originally proposed for the high-$T_c
$ cuprate.

Very few exact results are known for the $t$-$J$ model\cite{Leung1995,Hasegawa1989,Sarkar1990, Zheng2018},  but it has been
rigorously established that the statistical fermion sign structure in a
weakly-interacting electron system will be replaced by the statistical
phase-string sign structure in the bipartite $t$-$J$ model at any doping, temperature, and
dimensions\cite{Sheng1996, Weng1997, Wu2008,Weng2011, Wang2014, Zhu2016}. Previously, by DMRG\cite{Zhu2014Nature, Zhu2018} and variational Monte Carlo\cite{Chen2018} (VMC) studies, it has
been also revealed that the phase-string sign structure plays a critical
role in the pairing of two holes in the
two- and four-leg $t$-$J$ ladders to indicate that the pairing mechanism of RVB-type%
\cite{FRADKIN1990, Lee2006} is not sufficient.

These motivate us to systematically study the ground state of the two-leg $t$%
-$J$ ladder at finite doping by DMRG. In this paper, we shall show that it is
indeed an LE liquid similar to the four-leg case at finite doping concentration.\cite{Jiang2018tj} The
single-particle Green's function and the spin-spin correlation clearly show
exponential decays indicating a gap opening in the single particle channel
due to forming Cooper-pairing. However, such an LE state can be reduced to
a non-pairing Luttinger liquid (LL)\cite{Haldane1994}, very close to the \emph{%
free} Fermi gas limit, as soon as the phase-string sign structure is turned
off either by inserting a spin-dependent sign to the hopping term or by
making the charge-spin recombination in a strongly anisotropic case. By
further making an adiabatic continuation of the LE state in the limit of rung
hopping $t_{\perp}\rightarrow 0$, a hidden stringlike pairing force, originating from the phase-string sign
structure, can be revealed. It resembles an ``Amperean-like pairing'', \cite {Lee2007,Lee2014,Chen2019} and is
predominantly responsible for the strong pairing in the LE ground state that goes beyond the conventional RVB mechanism.

\emph{Model Hamiltonian.}---The hole-doped $t$-$J$ model on a square lattice
is defined by $H_{\mathrm{t\text{-}J}} = H_t + H_J$, with 
\begin{eqnarray}  \label{Eq:tJ_model}
H_t &=& - \sum_{\langle ij\rangle \sigma} t_{ij} \left(\hat{c}%
^{\dagger}_{i\sigma} \hat{c}_{j\sigma} + h.c.\right), \notag \\
H_J &=& \sum_{\langle ij\rangle} J_{ij} \left (\hat{\mathbf{S}}_i\cdot \hat{%
\mathbf{S}}_j - \frac{\hat{n}_i \hat{n}_j}{4} \right),
\end{eqnarray}
where $\hat{c}^\dagger_{i\sigma}$ ($\hat{c}_{i\sigma}$) is the electron
creation (annihilation) operator on site $i=(x_i,y_i)$ with spin $\sigma$. $%
\hat{\mathbf{S}}_i$ is the spin operator and $\hat{n}_i=\sum_{\sigma}\hat{c}%
^\dagger_{i\sigma}\hat{c}_{i\sigma}^{}$ is the electron number operator. $%
\langle ij\rangle$ denotes nearest-neighbor (NN) sites and the Hilbert space
is constrained by the no-double occupancy condition $\hat{n}_i\leq 1$. $%
t_{ij}$ is the electron hopping integral and $J_{ij}$ is the spin exchange
interaction between NN sites. Specifically, $t_{ij}=t_\parallel$ and $%
J_{ij}=J_\parallel$ for the intra-chain couplings, and $t_{ij}=t_\perp$ and $%
J_{ij}=J_\perp$ for the inter-chain couplings. For comparison, we shall also study the so-called $\sigma\cdot$$ t$-$J$ model $H_{%
\mathrm{\sigma}\cdot t\text{-}J} = H_{\sigma\cdot t} + H_J$,\cite{Zhu2012}
with the kinetic energy term $H_t$ in Eq.~(\ref{Eq:tJ_model}) replaced by%
\begin{eqnarray}  \label{Eq:stJ_model}
H_{\sigma \cdot t}= - \sum_{\langle ij\rangle \sigma} \sigma t_{ij} \left(%
\hat{c}^\dagger_{i\sigma} \hat{c}_{j\sigma} + h.c.\right),
\end{eqnarray}
where an extra spin-dependent sign $\sigma=\pm 1$ is added. It can be proven
that the phase-string sign structure hidden in the $t$-$J$ model is
precisely removed in the $\sigma\cdot$$ t$-$J$ model \cite{Zhu2012} (cf. Supplemental Material). Both models reduced to the same
antiferromagnetic (AFM) Heisenberg model at half-filling and the difference in sign structure only shows up upon doping.


\begin{table*}[tbp] \centering%

\caption{Summary of the phases: ``LE" (``LL") denotes the Luther-Emery (Luttinger) liquid. Corresponding central charge $c$ and exponents ($ K_c$, $K_G $, $K_{sc}$ and $K_s$) of the power-law behavior in the CDW amplitude $A_{cdw}$, single particle Green's function $G_\sigma$, pair-field correlation function $\Phi$ and spin-spin correlation function $F$ are shown (otherwise for an exponential decay, an length scale is marked by $\xi_G$ or $\xi_{s}$. 
Note that for $F$ of the $\sigma\cdot $$ t$-$J$ model, only the exponent for the $S_z$ component is shown, see text.)}
\label{Table I}%

\begin{tabular}{c|ccccccc}
\hline\hline
& Parameters  & Phase & $c$ & $A_{cdw}$ & $G_{\sigma }$ & $\Phi$ & $F$ \\ 
\hline
& $t_{\perp }=t_{\Vert },J_{\Vert }=J_{\perp }$ & LE & $1.27(3)$ & $1.06\left(
1\right) $ & $\xi _{G}\sim 5$ & $0.85\left( 2\right) $ & $\xi _{s}\sim 6$ \\ 
$t$-$J$ & $t_{\perp }=\gamma t_{\Vert },J_{\Vert }=J_{\perp }$ & LE & $%
1.27\sim 1.30$ &  &  &  &  \\ 
& $t_{\Vert }=0.4t_{\perp },J_{\Vert }=0.4J_{\perp }$ & LL & $2.08\left(
9\right) $ & $1.90\left( 2\right) $ & $~\sim 1$ & $2.12\left( 1\right) $ & $%
\sim 1.6$ \\ 
\hline
$\sigma \cdot $$t$-$J$ & $t_{\perp }=t_{\Vert },J_{\Vert }=J_{\perp }$ & LL & $%
2.04\left( 1\right) $ & $1.99\left( 1\right) $ & $1.10\left( 1\right) $ & $%
2.82\left( 1\right) $ & $1.93\left( 1\right) $ 
\end{tabular}%

\end{table*}%

We will focus on the following cases in this work: (1) Isotropic case with $%
t_\parallel=t_\perp=t$ and $J_\parallel=J_\perp=J$;
(2) Anisotropic in hopping: $t_\parallel=t$ and $%
t_\perp=\gamma t$, while $J_\parallel=J_\perp=J$, where $\gamma$ is a tuning
parameter; (3) Anisotropic in both hopping and superexchange terms: $%
t_\perp=t$ and $J_\perp=J$, while $t_\parallel=\alpha t$ and $%
J_\parallel=\alpha J$, where $\alpha$ is another tuning parameter. Here the
system is a square lattice two-leg ladder with system size $N=L_x\times L_y$%
, where $L_x$ and $L_y$ ($=2$) are the number of sites along the $\hat{x}%
=(1,0)$ and $\hat{y}=(0,1)$ directions, respectively, with the length up to $L_x=192$. The doping concentration $\delta=
\frac{N_h}{N}$, where $N_h$ is the number of doped holes measured from half-filling. 
Our calculation will mainly focus on a typical doping $\delta=12.5\%$ without loss of generality. We set $J=1$ as an energy unit and
consider $t=3$, and keep up to $m=8000$ number of states in each DMRG block
with truncation error $\epsilon\lesssim 5\times 10^{-9}$ and perform up to $%
60$ sweeps. This leads to excellent convergence for our results when extrapolated to $m=\infty$ limit.

\begin{figure}[tbp]
\includegraphics[width=\linewidth]{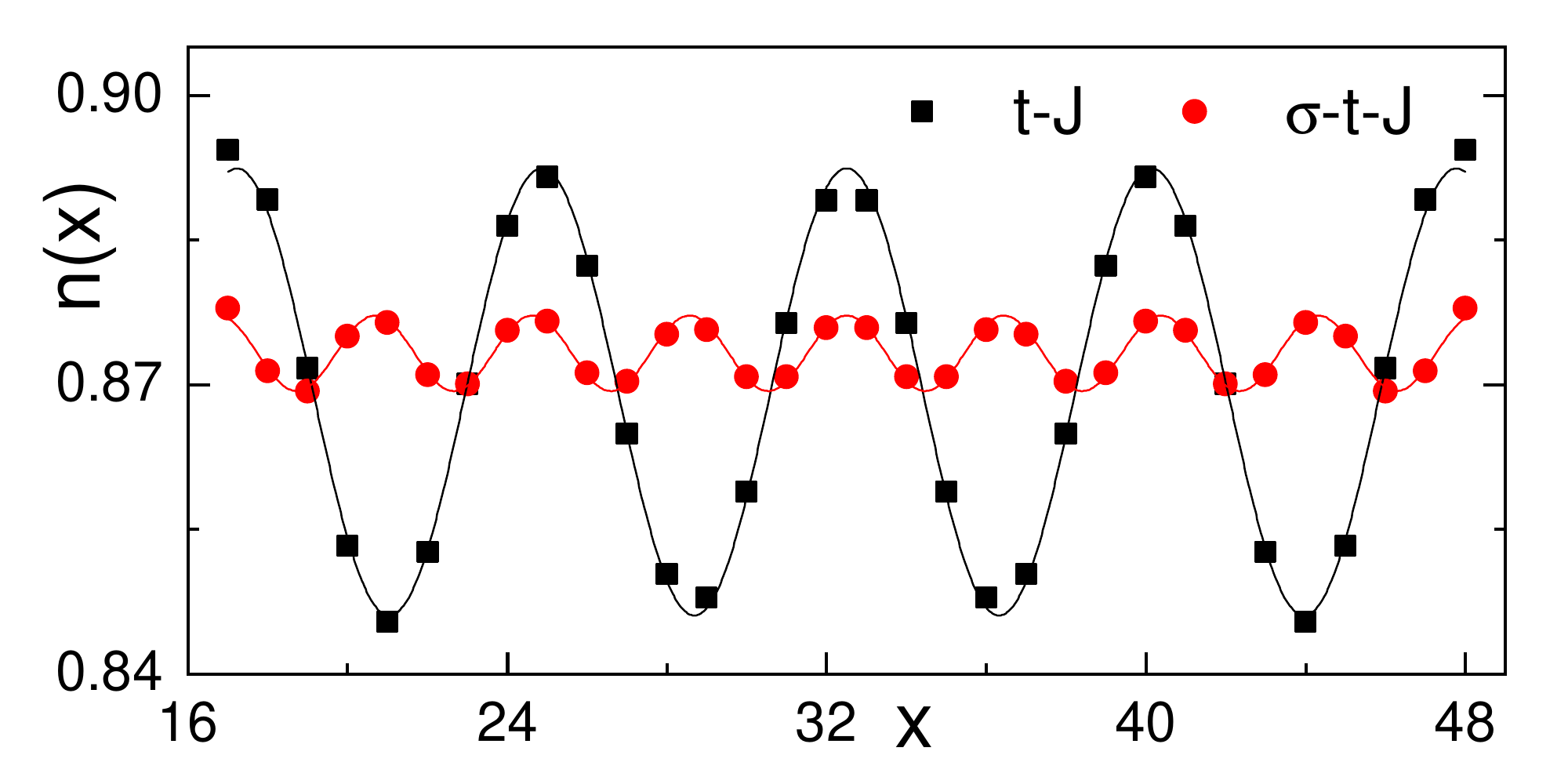} 
\caption{(Color online) Charge density profile $n(x)$ of the isotropic $t$-$J
$ and $\protect\sigma\cdot $$ t$-$J$ models at $\protect\delta=0.125$ and $L_x=64$, where only the central-half region are used to extract $A_{cdw}$ using Eq.(\ref{Eq:CDW}) labelled by the solid lines.}
\label{Fig:DensityProfile}
\end{figure}

\emph{Physical quantities.}---The following physical
quantities will be calculated by DMRG.\cite{White1992} The charge density $%
n(x)=\frac{1}{2}\sum_{y=1}^{2}\langle \hat{n}(x,y) \rangle$. The charge density wave (CDW) amplitude, $A_{cdw}$, inferred from $%
n(x)$ by 
\begin{eqnarray}  \label{Eq:CDW}
n(x)=A_{cdw}(L_x) \cos(Q_{cdw} x + \theta) + n_0~,
\end{eqnarray}
where $Q_{cdw}$ denotes the CDW wavevector, while $\theta$ and $n_0$ are fitting parameters. Since the ends of a
finite system break the translational symmetry, only the central-half region
with rung indices $\frac{L_x}{4}<x\leq\frac{3L_x}{4}$ is used in the fitting 
to minimize the boundary effect, as shown in Fig. \ref{Fig:DensityProfile} for $L_x$=$64$. 

The single-particle Green's function is defined as%
\begin{eqnarray}  \label{Eq:GF}
G_\sigma(r)=\frac{1}{2}\sum_{y=1}^{2} \langle \hat{c}^\dagger_{x_0,y,%
\sigma} \hat{c}_{x_0+r,y, \sigma}\rangle~,
\end{eqnarray}
where ($x_0,y$) is the reference site and $r$ is the displacement along the $%
\hat{x}$=(1,0) direction. If $G_\sigma(r)$ is short-ranged, it is
characterized by a length scale $\xi_G$: $G_\sigma(r)\sim e^{-r/\xi_G}$.
Otherwise, it is described by the Luttinger exponent $K_G$ in a power-law
behavior: $G_\sigma(r)\sim r^{-K_G}$.

A diagnostic of the SC order is by the pair-field correlation function 
\begin{eqnarray}
\Phi_{\alpha\beta}(r)=\frac{1}{2}\sum_{y=1}^{2}\ \langle
\Delta_\alpha^\dagger (x_0,y)\Delta_\beta(x_0+r,y)\rangle ~.  \label{Eq:SCOR}
\end{eqnarray}
Here the spin-singlet pair-field creation operator $\Delta_\alpha^%
\dagger(x,y)=\frac{1}{\sqrt{2}}[c_{(x,y),\uparrow}^\dagger
c_{(x,y)+\alpha,\downarrow}^\dagger - c_{(x,y),\downarrow}^\dagger
c_{(x,y)+\alpha,\uparrow}^\dagger]$, where bond orientations are designated $%
\alpha=\hat{x}$, $\hat{y}$, and $(x_0,y)$ is the reference bond and $r$ the
displacement along the $\hat{x}=(1,0)$ direction. Similarly, the spin-spin
correlation function is given by 
\begin{eqnarray}  \label{Eq:SpinCor}
F(r)=\frac{1}{2}\sum_{y=1}^{2}|\langle \hat{\mathbf{S}}_{x_0,y}\cdot 
\hat{\mathbf{S}}_{x_0+r,y}\rangle|~,
\end{eqnarray}
where $\hat{\mathbf{S}}_{x,y}$ denotes the spin operator at site $i=(x,y)$.
A spin gapped or gapless state is characterized by a short-ranged, $F(r)
\sim e^{-r/\xi_s}$, or quasi-long-ranged, $F(r) \sim r^{-K_s}$, respectively.

Finally, in the DMRG simulation, the central charge can be obtained by
calculating the von Neumann entropy $S=-\mathrm{Tr \rho ln \rho}$, where $%
\rho$ is the reduced density matrix of a subsystem with length $l$. For a
critical system, it has been established \cite{Calabrese2004} that $S(l)=%
\frac{c}{6}\mathrm{ln} (l)+\tilde{c}$ for open systems, where $c$ is the
central charge of the conformal field theory (CFT) and $\tilde{c}$ denotes a
model dependent constant. For finite cylinders with length $L_x$, we may fix 
$l=\frac{L_x}{2}$ and use the formula 
$S(\frac{L_x}{2})=\frac{c}{6}\mathrm{ln}(\frac{L_x}{2})+\tilde{c}$ 
to extract the central charge $c$, i.e., the number of gapless modes.

Table~\ref{Table I} summarizes the main results obtained by DMRG and the details will be discussed in the following.

\begin{figure}[tbp]
\includegraphics[width=\linewidth]{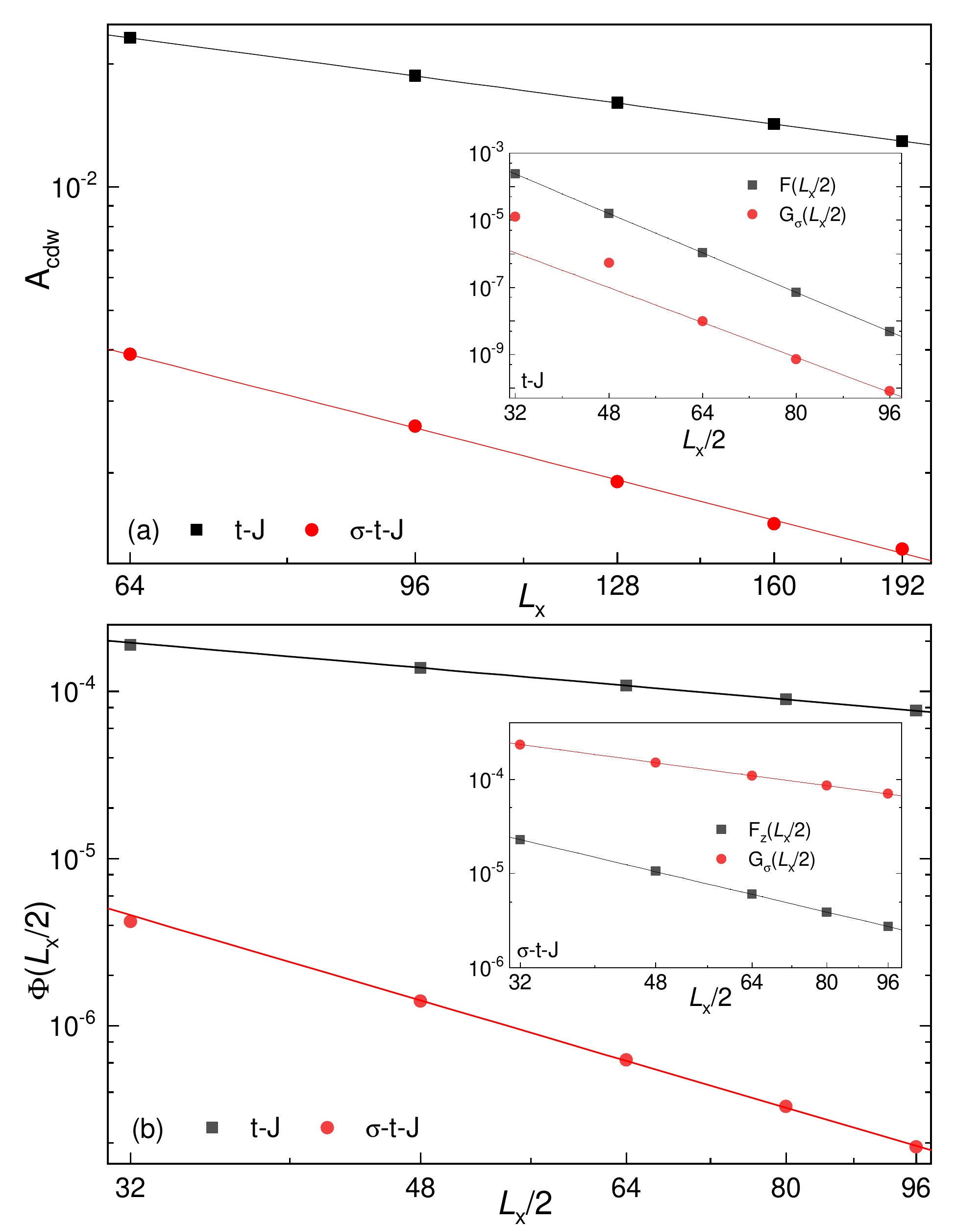} 
\caption{(Color online) Finite-size scalings of (a) CDW amplitude and (b) superconducting pair-field correlator 
for the isotropic $t$-$J$ and $\protect\sigma\cdot$$t$-$J$ models, respectively. Insets: Finite-size
scaling of the spin-spin and single-particle correlators for (a) $t$-$J$ model in semi-logarithmic
scales and (b) $\protect\sigma\cdot$$t$-$J$ model in double-logarithmic scales, respectively.}
\label{Fig:2}
\end{figure}

\emph{Superconducting/Luther-Emery state.}---The ground state of the two-leg $t$-$J$ ladder shows a
typical LE liquid behavior characterized by the following correlators (cf. Table~\ref{Table I}): 
\begin{eqnarray}
\Phi_{\alpha\beta}(r)&\propto & r^{-K_{sc}}, \\
A_{cdw}(L_x)&\propto& L_x^{-K_c/2}, \\
K_{c}{K_{sc}}&\sim &1 . 
\end{eqnarray}
In the isotropic case, one obtains $K_{sc}=0.85(2)$ and $K_c=1.06(1)$, respectively, for the SC pairing
and CDW amplitude, with $K_c K_{sc}\sim 1$ within the numerical error and
finite size effect (cf. Fig. \ref{Fig:2}). The density oscillates with a well-defined wavevector $%
Q_{cdw}$ with a wavelength $\lambda=\frac{1}{\delta}$ as shown in Fig.~\ref {Fig:DensityProfile}. 
The single-particle correlator has a length scale $\xi_G\sim 5$, while the spin-spin correlation length $\xi_s\sim 6$ (Fig. 
~\ref {Fig:2}): 
\begin{eqnarray}
G_\sigma({r})&\sim& e^{-r/\xi_G} , \\
F(r)&\sim & e^{-r/\xi_s} .
\end{eqnarray}
Moreover, the central charge extracted from the scaling of the entanglement
entropy, is $c=1.27(3)$, which is qualitatively consistent with one gapless
charge mode with $c=1$ in the LE liquid. The above LE liquid is similar to that of the four-leg $t$-$J$ ladder \cite{Jiang2018tj} and persists over a wide range of finite doping.

\begin{figure}[tbp]
\includegraphics[width=\linewidth]{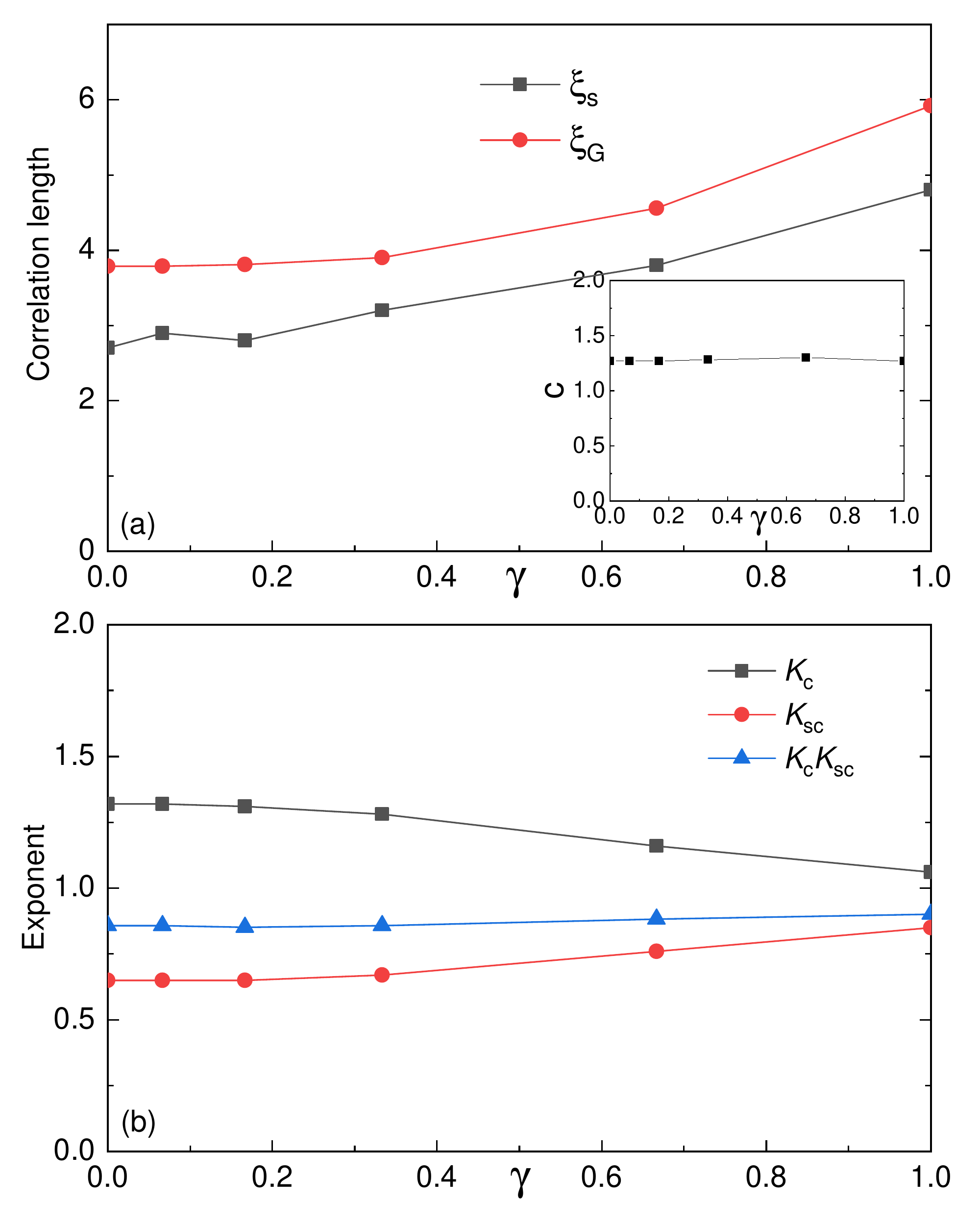}  
\caption{(Color online) The stable LE state in an anisotropic $t$-$J$ ladder with $\protect\gamma\equiv
t_\perp/t_\parallel$ and $J_\perp=J_\parallel=J$: (a) Correlation lengths and (b) the exponents of the power-law correlations. Inset in (a) shows the
central charge $c$.}
\label{Fig:tJ_tr}
\end{figure}

To show the robustness of the LE phase, we further examine the anisotropy in
the hopping: $t_\parallel=t$ and $t_\perp=\gamma t$, while the superexchange
coupling remains isotropic. From $\gamma=1\rightarrow 0$, we find that the ground state always remains an LE liquid with
the same modulation wavelengths of the charge density and spin density, and the product $K_c K_{sc}\sim 1$ in the whole regime of $0\leq \gamma \leq1$ [cf. Fig.~\ref{Fig:tJ_tr}(b)]. The correlation lengths, $\xi_G$ and $\xi_s$%
, slightly decrease with reducing $\gamma$ as shown in Fig.~\ref{Fig:tJ_tr}
(a). In the inset of Fig.~\ref{Fig:tJ_tr}(a), the central charge $c=1.27\sim
1.30$ is qualitatively consistent with one gapless charge mode with $c=1$.


\emph{Fermi liquid/Luttinger liquid state: Disappearance of the phase string sign structure.}%
---In sharp contrast, in the two-leg $\sigma\cdot$$t$-$J$ model at the same
doping, the ground state is qualitatively changed to an LL state characterized by (cf. Fig. ~\ref%
{Fig:2}) 
\begin{eqnarray}
G_\sigma({r})&\sim& r^{-K_G} , \\
F(r)&\sim& r^{-K_s},
\end {eqnarray}
with the dominant Luttinger exponent $K_G=1.10(1)$, very close to the Fermi
gas limit, while both the density-density and SC correlators become sub-leading,
with $K_c=1.99(1)$ and $K_{sc}=2.82(1)$, respectively. The
spin-spin correlation also changes to a power-law behavior with $K_s^{zz}=1.93(1)$ and 
$K_s^{xx,yy}=0.43(1)$, respectively (note the absence of spin rotation
symmetry in the hopping term of the $\sigma\cdot$$t$-$J$ model, see below).
Correspondingly the central charge $c=2.04(1)$ close to $c=2$. Here the charge modulation
wavelength becomes $\lambda=\frac{1}{2\delta}$ and spin modulation
wavelength $\frac{1}{\delta}$.


\begin{figure}[tbp]
\includegraphics[width=\linewidth]{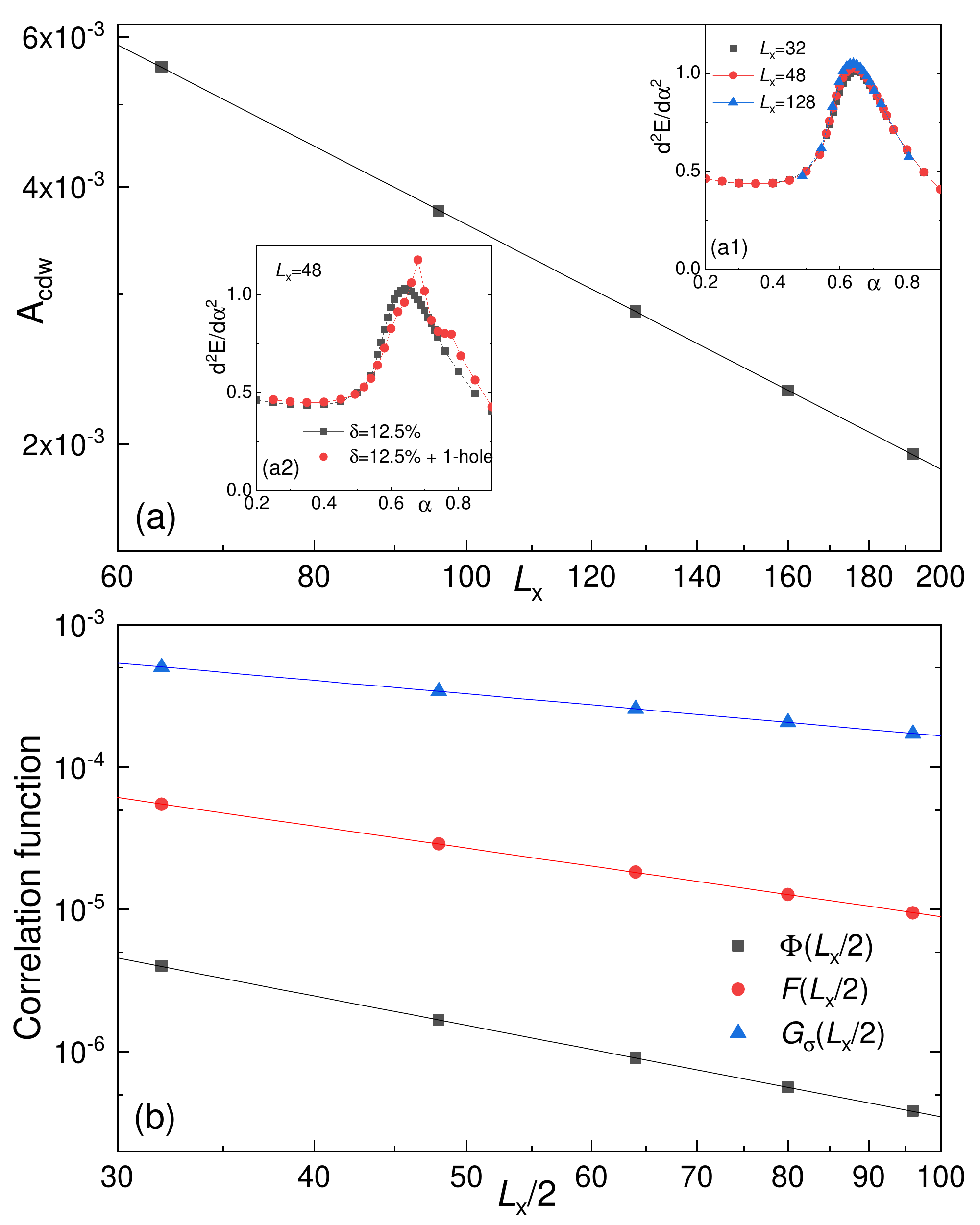} 
\caption{(Color online) Fermi-gas-like state in the strong rung regime ($\protect\alpha=0.4$) of the anisotropic $t$-$J$ ladder with $\protect \alpha \equiv t_\parallel/t_\perp \equiv J_\parallel/J_\perp$ and $%
t_\parallel=t$ and $J_\parallel=J$.  Finite-size scalings of (a) CDW amplitude and (b) superconducting pair-field,
spin-spin, and single-particle correlators. Insets are the second order
derivatives of the ground state energy density at doping $\delta=0.125$ (and with one more hole which determines $\protect\alpha_c\sim 0.68$ at $t/J=3$) (see text).}
\label{Fig:tJ_tcJc}
\end{figure}

Furthermore, similar Fermi gas/LL ground state is also identified even in the $t$-$J
$ ladder in a strong rung coupling case. With $t_\perp=t$ and $J_\perp=J$, while $%
t_\parallel=\alpha t$ and $J_\parallel=\alpha J$, a transition to an LL state is found at $\alpha<\alpha_c\sim 0.68$ (for $t/J=3$) in the $t$-$J$ ladder. For example, as shown in Fig.~\ref{Fig:tJ_tcJc} at $\alpha=0.4$, the Luttinger exponent $K_G\sim 1$, while the decay of the CDW amplitude $A_{cdw}(L_x)$ with length $L_x$ and the
pair-field correlator $\Phi(r)$ at large distance also become sub-leading: $K_c=1.90(2)$ and $K_{sc}=2.12(1)$%
, respectively, with the central charge $c=2.08(9)$, which are all similar to the $\sigma\cdot$$ t$-$J$ case. However, it is noted that the spin SU(2) rotation symmetry is still maintained here and the spin-spin
correlator is specified by a single exponent $K_s\sim 1.6$. 

As a matter of fact, a transition to a conventional quasiparticle state has been previously identified in the same anisotropic $t$-$J$ ladder in the strong rung coupling at $\alpha_c\sim 0.68$ for the single-hole-doped case. \cite{Zhu2012,Zhu2015,Zhu2015Quasiparticle,White2015} There it has been shown \cite{Zhu2015,Zhu2015Quasiparticle} that due to the strong recombination of the hole and its spin partner, the phase string sign structure is indeed effectively removed in the strong rung limit $\alpha<\alpha_c$. In the present finite doping at $\delta=0.125$, the ground state at $\alpha<\alpha_c$ still remains in a Fermi gas state, with $\alpha_c\sim 0.68$ essentially independent of doping. Here $\alpha_c$ is determined by the second-order derivative of the ground state energy density as shown in the inset (a2) of Fig.~\ref{Fig:tJ_tcJc}(a) where one more hole is added on top of the all paired $\delta=12.5\%$ holes. For comparison, a much smoother peak at $\alpha_c\sim 0.64$ is also shown in the second-order derivative of the energy for the paired ground state at $\delta=12.5\%$ holes in the insets of Fig.~\ref{Fig:tJ_tcJc}(a). Such an ``SC'' transition point coincides with the critical point for binding between two doped holes, \cite{Zhu2014Nature} which is also slightly lowered than $\alpha_c$ for the single-hole case.

\emph{Non-BCS nature of pairing in the $t$-$J$ model.}---We have found that the
LE liquid as a prototypical SC phase in the quasi-1D system can make a
transition to a non-pairing LL phase by switching off the phase-string sign
structure either in the $\sigma\cdot$$t$-$J$ model or at $\alpha<\alpha_c$
in an anisotropic $t$-$J$ ladder. In the following, let us explicitly show
how such a novel sign structure plays a critical role in the pairing of the
LE state.

First, let us recall that the LE state remains smooth as a function of $%
\gamma $ and persists in the limit of $\gamma \rightarrow 0$ as the hopping $%
t_{\perp }$ diminishes while the $J$-term remains isotropic in the $t$-$J$
model (cf. Fig. ~\ref{Fig:tJ_tr}). In this limit, a duality transformation $%
e^{i\hat{\Theta}}$ can be explicitly performed to turn the $t$-$J$ model
into the phase-string-free $\sigma \cdot t$-$J$ model plus an additional
\textquotedblleft stringlike\textquotedblright\ pairing term as previously
shown for the two-hole case, which still holds true in an arbitrary
many-hole case\cite{Zhu2018, Chen2018}, 
\begin{equation}
{H}_{\mathrm{t\text{-}J}}\rightarrow \widetilde{H}_{\mathrm{t\text{-}J}}=H_{%
\mathrm{\sigma \cdot t\text{-}J}}+H_{\mathrm{I}}^{\mathrm{string}}
\label{tildeH}
\end{equation}%
where 
\begin{equation}
H_{\mathrm{I}}^{\mathrm{string}}=-\frac{1}{2}J\sum_{x_i}(\hat{S}%
_{(x_{i},1)}^{+}\hat{S}_{(x_{i},2)}^{-}+\hat{S}_{(x_{i},1)}^{-}\hat{S}%
_{(x_{i},2)}^{+})(1-\Delta \Lambda _{i}^{h})  \label{HI}
\end{equation}%
in which the summation over $i$ is along the chain direction, and 
$\Delta \Lambda _{i}^{h}=\exp \left[ {-i\pi \sum\limits_{x_{l}<x_{i}}(\hat{n}%
_{\left( x_{l},1\right) }^{h}-\hat{n}_{\left( x_{l},2\right) }^{h})}\right] $
describes the nonlocal phase shift effect created by the doped holes at both
chains (legs). Since the transverse spin at each rung: $\langle \hat{S}%
_{(x_{i},1)}^{+}\hat{S}_{(x_{i},2)}^{-}+\hat{S}_{(x_{i},1)}^{-}\hat{S}%
_{(x_{i},2)}^{+}\rangle <0$ as ensured by $J$ at half-filling and finite
doping, one finds that doped holes will generally acquire a string-like 
\emph{strong} pairing potential in Eq. (\ref{HI}), 
in addition to the usual $J$-term in $H_{\mathrm{\sigma \cdot t\text{-}J}}$. It
will then result in an strong pairing ground state $|\Psi _{\mathrm{BCS}%
}\rangle $ for $\tilde{H}_{\mathrm{t-J}}$.%

The true LE ground state of the original ${H}_{\mathrm{t\text{-}J}}$ (at $t_{\perp
}=0$) is then written by 
\begin{equation}
|\Psi _{\mathrm{G}}\rangle =e^{i\hat{\Theta}}|\Psi _{\mathrm{BCS}}\rangle ~,
\label{unitary}
\end{equation}%
where the duality transformation 
$\hat{\Theta}\equiv -\sum\limits_{i}\hat{n}_{\left( x_{i},y_{i}\right) }^{h}%
\hat{\Omega}_{i}$ 
and $\hat{\Omega}_{i}=\pm \pi \sum_{y=1}^{L_{y}}\sum_{x_{l}>x_{i}}n_{\left(
x_{l},y\right) }^{\downarrow }$, where $n_{\left( x_{l},y\right)
}^{\downarrow }$ is the number operator of down spin at site $\left(
x_{l},y\right) $. Therefore, the phase string sign structure as represented
by $e^{i\hat{\Theta}}$ is topological and non-perturbative, which gives rise
to a non-BCS form [Eq.~(\ref{unitary})] of the ground state with an
Amperean-like novel pairing force shown in Eq. (\ref{tildeH}). Alternatively in the
supplemental material, a bosonization method has been applied to treat
the LE ground state in the large $\beta \equiv
J_{\perp }/J_{\parallel }$ limit.

\textbf{Acknowledgement:} We would like to thank S. Kivelson and Z. Zhu for insightful discussions. H.C.J. was supported by the Department of Energy, Office of Science, Basic Energy Sciences, Materials Sciences and Engineering Division, under Contract DE-AC02-76SF00515; S.C. and Z.W. are partially supported by Natural Science Foundation of China (Grant No. 11534007), MOST of China (Grant Nos. 2015CB921000 and 2017YFA0302902). Parts of the computing for this project was performed on the Sherlock cluster.


\newcommand\TODO[1]{\textcolor{red}{#1}}
\renewcommand{\thefigure}{S\arabic{figure}}
\setcounter{figure}{0}
\renewcommand{\theequation}{S\arabic{equation}}
\setcounter{equation}{0}
\renewcommand{\thesection}{\arabic{section}}

\newpage

\section{Supplementary Material} %
This supplementary material contains two parts. In the first part, we outline the rigorous \emph{phase string} sign structures in a bipartite $t$-$J$ model and the absence of this novel statistical sign structure in the $\sigma\cdot$$t$-$J$ model, and discuss the implication for the comparative study in the main text. In the second part, we provide an analytic study of the Luther-Emery liquid ground state for the $t$-$J$ two-leg ladder in the limit of $t_{\perp}=0$ and $J_{\perp}\gg J_{\parallel }$.

\section{Sign structure}

The $t$-$J$ and $\sigma \cdot $$t$-$J$ models with $H_{t\text{-}J}=H_{t}+H_{J}$ and $H_{\sigma
\cdot t\text{-}J}=H_{\sigma \cdot t}+H_{J}$ 
\begin{eqnarray}\label{H}
H_{t} &=&-\sum_{\left\langle ij\right\rangle \sigma }t_{ij}\hat{c}_{i\sigma
}^{\dag }\hat{c}_{j\sigma }+\text{h.c}, \\
H_{\sigma \cdot t} &=&-\sum_{\left\langle ij\right\rangle \sigma }t_{ij}\hat{%
c}_{i\sigma }^{\dag }\hat{c}_{j\sigma }+\text{h.c}, \\
H_{J} &=&\sum_{\left\langle ij\right\rangle }J_{ij}\left( \mathbf{\hat{S}}%
_{i}\cdot \mathbf{\hat{S}}_{j}-\frac{1}{4}\hat{n}_{i}\hat{n}_{j}\right)
\end{eqnarray}%
The many-body Hilbert space is subject to a no-double occupancy constraint 
\begin{equation}
\sum_{i}\hat{n}_{i}\leq 1.  \label{app:constraint}
\end{equation}%
Here $\hat{c}_{i\sigma }$ annihilates an electron with spin $\sigma $ at
site $i$. And $\mathbf{\hat{S}}_{i}$ and $\hat{n}_{i}$ are spin and electron
number operators respectively at site $i$. Specifically, $%
t_{ij}=t_{\parallel }$ and $J_{ij}=J_{\parallel }$ for the intra-chain
couplings and $t_{ij}=t_{\perp }$ and $J_{ij}=J_{\perp }$ fro the
inter-chain couplings.


The $t$-$J$ model is considered to be one of the simplest model to describe the spin full doped Mott insulator. The strong correlation nature of the Mott physics may be well represented by the novel statistical sign structure hidden in the $t$-$J$ model, which has been demonstrated in Ref.~%
\onlinecite{Sheng1996, Weng1997, Wu2008}. For a bipartite lattice of any dimensions, doping concentration, and temperature, the partition function of the $t$-$J$ model can be generally expressed as \cite{Wu2008}
\begin{equation}
Z_{t\text{-}J}=\sum_{c}\tau _{c}\mathcal{Z}\left[ c\right] 
\label{tjpartition}
\end{equation}%
with each path $c$ composed of a set of closed loops of the spatial
trajectories of all holes and $\mathcal{Z}[c]\geq 0$.\cite{Wu2008} The general sign factor $%
\tau _{c}$ in Eq. (\ref{tjpartition}) is given by
\begin{equation}\label{tau}
\tau _{c}\equiv (-1)^{N_{h}^{\downarrow }[c]+N_{\mathrm{ex}}^{h}[c]}~.
\end{equation}%
Here the Berry-phase-like sign factor $\left( -1\right) ^{N_{h}^{\downarrow
}[c]}$ is associated with the hopping processes of the exchanging between holes and
spin-$\downarrow$, which is known as the phase string enforced on each hole closed loop in Eq. (\ref{tjpartition}). Such a novel sign structure replaces the conventional fermion signs for the electrons and implies an intrinsic mutual statistics between holes and spins, which further suggests a new type of fractionalization \cite {Weng2011,Chen2019,Zhang2019}. Furthermore, there is another sign factor $\left( -1\right) ^{N_{\mathrm{ex}}^{\mathrm{h%
}}[c]}$ in Eq. (\ref{tau}), which resembles a conventional Fermi-Dirac statistical signs associated with hole-hole exchange process as identical particles. 

The novel phase-string sign in Eq.~(\ref{tau}) can be switched off
by inserting a sign $\sigma $ in each hopping term of $H_t$ to result in $H_{\sigma \cdot t}$ as given in Eq.(S2). Then one obtains a phase-string-free $\sigma \cdot $$t$-$J$ model whose partition function reduces to
\begin{equation}
Z_{\sigma \cdot t\text{-}J}=\sum_{c}\left( -1\right) ^{N_{\mathrm{ex}}^{%
\mathrm{h}}\left[ c\right] }\mathcal{Z}\left[ c\right] .
\label{sigmatjmodel}
\end{equation}
where the fermion signs betweeen holes are unchanged and $\mathcal{Z}[c]\geq 0$ remains the same as in Eq. (\ref{tjpartition}). But the phase string sign factor is precisely removed. 

To understand the two-dimensional (2D) $\sigma \cdot $$t$-$J$ model further, let us introduce a transformation 
\begin{eqnarray}
c_{\left( x,\nu \right) \uparrow } &\rightarrow &c_{\left( x,\nu \right)
\uparrow }, \\
c_{\left( x,\nu \right) \downarrow } &\rightarrow &\left( -1\right) ^{\nu
}\left( -1\right) ^{x}c_{\left( x,\nu \right) \downarrow },
\end{eqnarray}%
where each site $i=\left( x,\nu \right) $ is specified by two coordinates in 2D. 
Then the $\sigma \cdot $$t$-$J$ model is transformed into a 2D doping XXZ model:
\begin{equation}
H_{\sigma \cdot t\text{-}J}\rightarrow H_{\mathrm{dXXZ}}
\end{equation}%
where $H_{\mathrm{dXXZ}} =H_{t}+H_{\mathrm{XXZ}}$, with
\begin{eqnarray}
H_{\mathrm{XXZ}} =\sum_{\left\langle ij\right\rangle } J_{ij}(
S_{i}^{z}S_{j}^{z}-\frac{1}{4}\hat{n}_{i}\hat{n}_{j})-J_{ij}\left(S_{i}^{x}S_{j}^{x}+S_{i}^{y}S_{j}^{y}\right) .
\end{eqnarray}%
Compared with 
$H_{t\text{-}J}$, $H_{\mathrm{dXXZ}}$ now has an antiferromagnetic spin background
with a ferromagnetic interaction in the XY-plane. In other words, the doped holes now feel like as if they are moving in a much less frustrated quantum spin background where the spins in the easy (XY)-plane are in ferromagnetic array.  Of course, in the original $\sigma \cdot $$t$-$J$ model, the $J$-term still remains the same as in the $t$-$J$ model case. It is the hopping term that is changed to remove the phase string frustration.

Therefore, the essential Mott physics, which is normally associated with the no double occupancy constraint in Eq. (\ref{app:constraint}), is really given by the phase string sign structure of Eq. (\ref{tau}). What we have shown above is that the novel phase string effect can be precisely distinguished by the difference between the $t$-$J$ and $\sigma \cdot $$t$-$J$ models, which is shown in the DMRG study as given in the main text for the two-leg ladder case at finite doping.

Finally, we point out that in the one-dimensional (1D) case, the $t$-$J$ and $\sigma \cdot $$t$-$J$ models can be further connected by a unitary transformation. Namely, under the open boundary condition, one finds $H_{\sigma\cdot t\text{-}J}=\mathcal{U}_{\mathrm{1D}}H_{ t%
\text{-}J}\mathcal{U}_{\mathrm{1D}}^{\dag }$ by
\begin{equation}\label{1D transf}
\mathcal{U}_{\mathrm{1D}}=\prod_{l}\exp \left( -i\pi
\sum_{j>l}n_{l}^{h}n_{j}^{\downarrow }\right) 
\end{equation}%
where $n_{l}^{h}\equiv 1-n_{l}$ and $n_{l}^{\downarrow }$ are number operators of holes and down spin at site $l$. 
Similar transformation can be then constructed in a two-leg ladder system
with $t_{\perp }=0$ such that the hopping term reduces to the 1D like two-decoupled chains but the superexchanger $J_{\perp}$ remains finite in the rung of the ladder. It is similar to the 1D version of Eq. (\ref{1D transf}) as given by 
\begin{equation}
|\Psi _{\mathrm{t-J}}\rangle =e^{i\hat{\Theta}}|\tilde{\Psi} _{\mathrm{t-J}}\rangle ~,
\label{unitary}
\end{equation}%
where
\begin{equation}\label{duality}
\hat{\Theta}\equiv -\sum\limits_{i}\hat{n}_{\left( x_{i},y_{i}\right) }^{h}%
\hat{\Omega}_{i}
\end{equation}
where $\hat{\Omega}_{i}=\pm \pi \sum_{y=1}^{L_{y}}\sum_{x_{l}>x_{i}}n_{\left(
x_{l},y\right) }^{\downarrow }$, where $n_{\left( x_{l},y\right)
}^{\downarrow }$ is the number operator of down spin at site $\left(
x_{l},y\right) $.
Then it is a straightforward to obtain
\begin{equation}
{H}_{\mathrm{t\text{-}J}}\rightarrow H_{\mathrm{\sigma \cdot t\text{-}J}}+H_{\mathrm{I}}^{\mathrm{string}}
\label{tildeH}
\end{equation}%
where 
\begin{equation}
H_{\mathrm{I}}^{\mathrm{string}}=-\frac{1}{2}J_{\perp }\sum_{x_i}(\hat{S}%
_{(x_{i},1)}^{+}\hat{S}_{(x_{i},2)}^{-}+\hat{S}_{(x_{i},1)}^{-}\hat{S}%
_{(x_{i},2)}^{+})(1-\Delta \Lambda _{i}^{h})  \label{HI}
\end{equation}%
in which the summation over $i$ is along the chain direction, and 
$\Delta \Lambda _{i}^{h}=\exp \left[ {-i\pi \sum\limits_{x_{l}<x_{i}}(\hat{n}%
_{\left( x_{l},1\right) }^{h}-\hat{n}_{\left( x_{l},2\right) }^{h})}\right] $
describes the nonlocal phase shift effect created by the doped holes at both
chains (legs). 

It is easy to see that $H_{\mathrm{I}}^{\mathrm{string}}$ vanishes if $J_{\perp }=0$, but with $J_{\perp }\neq 0$ in a two-leg ladder coupling, a string-like or ``Amperean-like'' pairing force emerges in additional to the usual J term in $H_{\mathrm{\sigma \cdot t\text{-}J}}$. Since the transverse spin at each rung: $\langle \hat{S}%
_{(x_{i},1)}^{+}\hat{S}_{(x_{i},2)}^{-}+\hat{S}_{(x_{i},1)}^{-}\hat{S}%
_{(x_{i},2)}^{+}\rangle <0$ as ensured by $J_{\perp}\neq 0$, one finds that doped holes will generally acquire a string-like 
\emph{strong} pairing potential in Eq. (\ref{HI}), which
will then result in an strong pairing ground state $|\tilde{\Psi} _{\mathrm{t-J}}\rangle $ in the transformed $t$-$J$ ladder.

 
Alternatively, in the following we shall treat the strong string-like pairing potential in a straightforward way in the large $J_{\perp }/J_{\parallel }\gg 1$ limit, where a bosonization method can be applied to show that the ground state is indeed an LE liquid, consistent with the DMRG in the main text.


\section{The LE ground state for $H_{t_{\parallel}\text{-}J}$ at $J_{\perp }/J_{\parallel } \gg1$}

The ground state of the two-leg $t$-$J$ ladder stays stable from large $\gamma $ to $\gamma \rightarrow 0$ limit (cf. Fig. 3 in the main text) as an LE liquid. It stays smooth also at $\gamma = 0$ over a large range of $\beta \equiv J_{\perp }/J_{\parallel }$. For the following analytic analysis, we shall consider the limit $\beta\gg 1$ at finite doping case.

\subsection{Effective model}

\begin{figure}[th]
\includegraphics[scale=0.4]{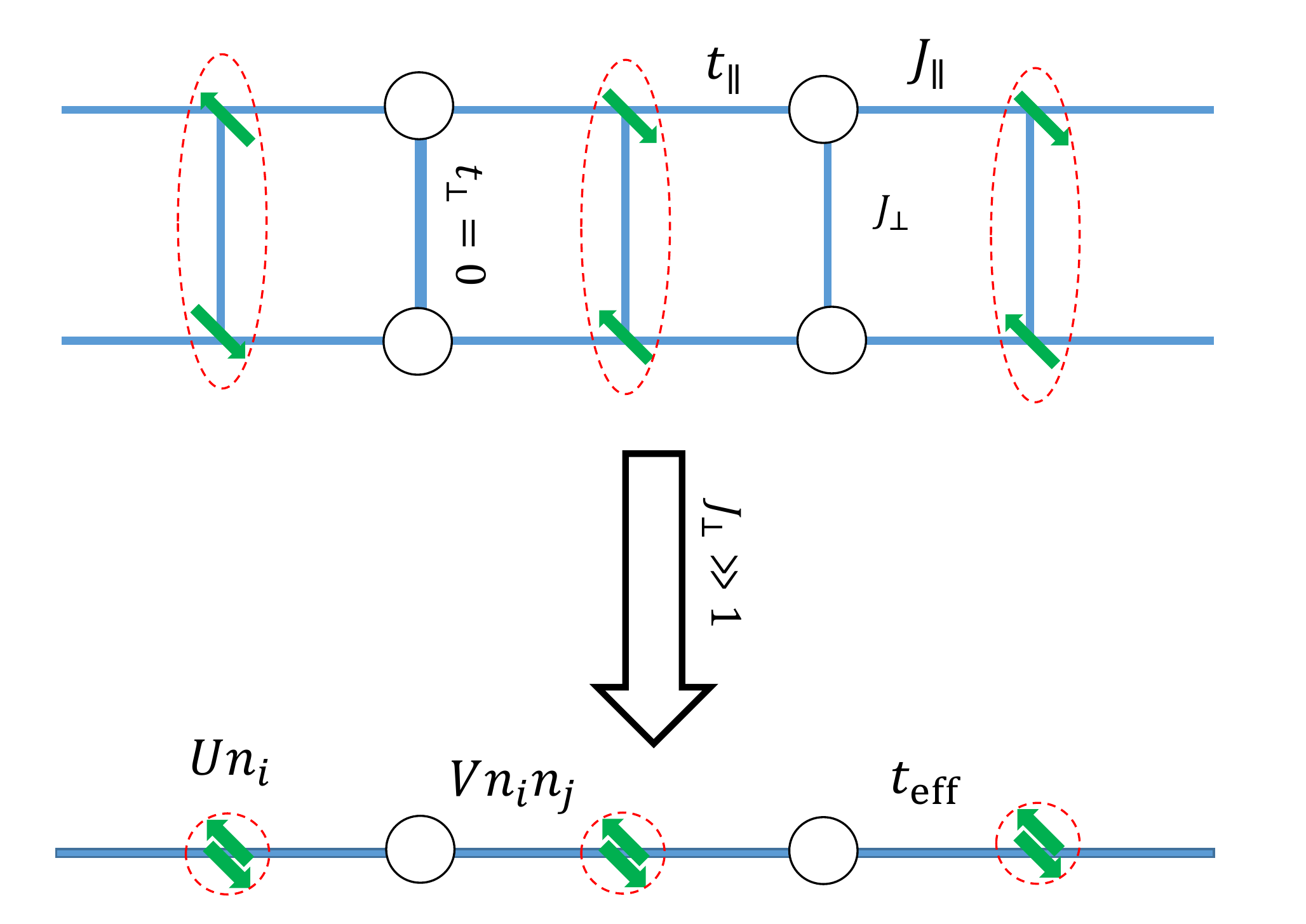}
\caption{A schematic illustration of an effective hard-core chain obtained by compressing the
two-leg $t$-$J$ ladder in the large $J_{\perp }/J_{\parallel }$ limit with interchain
hopping $t_{\perp }=0$.}
\label{app:FigS1}
\end{figure}
At $\beta\gg 1$, one has the following perturbative scheme for the $t$-$J$ ladder. Define 
\begin{equation}
H_{0}=H_{J_{\perp }}\mbox{, }H_{\text{int}}=H_{t_{\parallel
}}+H_{J_{\parallel }}
\end{equation}
then a large $\beta$ makes it possible to compress two legs into
a single chain (cf. Fig. \ref{app:FigS1}). More explicitly, with a Nakajima
transformation\cite{ZHONG1991}, we obtain the effective Hamiltonian $H_{%
\mathrm{eff}}$ as follows: 
\begin{eqnarray}
H_{\mathrm{eff}} &=&\sum_{i}-\frac{8}{3}\frac{t^{2}}{J_{\perp }}\left(
b_{i}^{\dag }b_{i+1}^{{}}+b_{i+1}^{\dag }b_{i}^{{}}\right) -\frac{%
J_{\parallel }^{2}}{4J_{\perp }}n_{i}n_{i+1}-\frac{3}{4}J_{\perp }n_{i} 
\nonumber \\
&=&\sum_{i}-t_{\text{eff}}\left( b_{i}^{\dag }b_{i+1}^{{}}+b_{i+1}^{\dag
}b_{i}^{{}}\right) -V\hat{n}_{i}^{b}\hat{n}_{i+1}^{b}-U\hat{n}_{i}^{b}
\label{app:Effmodel}
\end{eqnarray}%
where the operator $b_{i}$ annihilates a pair of electrons in the original
system 
\begin{equation}
b_{i}\Longleftrightarrow \sum_{\sigma }\sigma \hat{c}_{1i\sigma }\hat{c}%
_{2i-\sigma }
\end{equation}%
and $\hat{n}_{i}^{b}$ is the number of pairs on the rung $i$ that only can
take value $1$ or $0$. The effective model $H_{\mathrm{eff}}$ in Eq.~(\ref%
{app:Effmodel}) describes a 1D hard-core bose system with a nearest
neighbour attractive interaction. One can then solve it via the bosonization method.

We introduce the Jordan-Wagner transformation, 
\begin{equation}
\psi _{i}^{{}}=b_{i}^{{}}\exp \left( -i\pi \sum_{l>i}n_{i}\right) 
\end{equation}%
We obtain a fermionic form of Eq.~(\ref{app:Effmodel}) 
\begin{equation}
H_{\mathrm{eff}}=-t_{\text{eff}}\left( \psi _{i}^{\dag }\psi
_{i+1}^{{}}+\psi _{i+1}^{\dag }\psi _{i}^{{}}\right) -Vn_{i}n_{i+1}-Un_{i}
\end{equation}%
With the bosonized field, 
\begin{equation}
\psi _{i}=\frac{\eta }{\sqrt{2\pi a}}e^{ik_{b}x}e^{-i\left( \phi _{b}-\theta
_{b}\right) }+\frac{\bar{\eta}}{\sqrt{2\pi a}}e^{-ik_{b}x}e^{-i\left( -\phi
_{b}-\theta _{b}\right) }
\end{equation}%
with $\eta $ and $\bar{\eta}$ Klein factor obeying an anticommutation
relation, the corresponding bosonic Hamiltonian 
\begin{equation}
H_{\mathrm{eff}}=\frac{u_{b}}{2\pi }\int dx\left[ K_{b}\left( \partial
_{x}\theta _{b}\left( x\right) \right) ^{2}+\frac{1}{K_{b}}\left( \partial
_{x}\phi _{b}\left( x\right) \right) ^{2}\right] -\frac{U}{\pi }\partial
_{x}\phi _{b}
\end{equation}%
describes low energy fluctuations of the pairing field $\phi $ with the
stiffness constant%
\begin{equation}
K_{b}=\sqrt{\frac{u_{0}}{u_{0}-V\frac{2\sin ^{2}\left( k_{f}a\right) }{\pi }}%
}=1+V\frac{2\sin ^{2}\left( k_{b}a\right) }{\pi u_{0}}.
\end{equation}%
Here $K_{b}$ is the Luttinger parameter $k_{b}=\pi\left( 1-\delta
\right) $ is the effective fermi momentum. The constant potential term $%
\frac{U}{\pi }\partial _{x}\phi $ can be ignored.

\subsection{Luther-Emery Liquid}

\subsubsection{Density-density correlation}

\label{app:d-dcor}

The operators in the $t$-$J$ model should also be mapped into effective
operators in Eq.~(\ref{app:Effmodel}). For example, the hole number operator 
$n^{h}\left( x\right) \equiv \frac{1}{2}\sum_{\nu }\left( 1-n_{\left( x,\nu
\right) }\right) $ for $\hat{c}_{\left( x,\nu \right) }$ is mapped into the
pairing number operator $n_{x}^{b}$ for $b_{x}$. From the pair
density-density correlator $\left\langle n_{i}^{b}n_{j}^{b}\right\rangle $%
\begin{eqnarray}
\left\langle n^{h}\left( x\right) n^{h}\left( y\right) \right\rangle 
&=&\left\langle n_{x}^{b}n_{y}^{b}\right\rangle  \\
\! &=&-\frac{1}{\pi ^{2}}\frac{K_{b}}{r^{2}}+\frac{2}{\left( 2\pi \right)
^{2}}\left\vert r\right\vert ^{-2K_{b}}\cos \left( 2k_{f}r\right) 
\end{eqnarray}%
with the formula for the density operator 
\begin{equation}
n_{x}^{b}=-\frac{1}{\pi }\partial _{x}\phi _{b}+\frac{1}{2\pi }\left[
e^{i2k_{b}x}e^{-2i\phi _{b}}+\text{h.c.}\right] ~.
\end{equation}%
We can deduce $K_{c}=2K_{b}$.

\subsubsection{Superconducting pairing correlation \label{app:sccor}}

The singlet superconducting pairing correlation $\left\langle
b_{y}b_{x}^{\dagger }\right\rangle $ can be described in the fermionic representation by
(suppose $x>y$) 
\begin{eqnarray}
\left\langle b_{y}b_{x}^{\dagger }\right\rangle &=&\left\langle \exp \left(
-i\pi \sum_{l>y}\hat{n}_{l}\right) \psi _{y}\psi _{x}^{\dag }\exp \left(
+i\pi \sum_{l>x}\hat{n}_{l}\right) \right\rangle  \nonumber \\
&\sim &e^{ik_{f}r}|r|^{-\frac{1}{2}K_{b}^{-1}} 
\end{eqnarray}
such that $K_{sc}=\frac{1}{2}K_{b}^{-1}.$

\bigskip

Finally we obtain an important relation for $K_c$ and $K_{sc}$ in Sec \ref{app:d-dcor}
and Sec \ref{app:sccor} of the main text:
\begin{equation}
K_{c}K_{sc}=1
\end{equation}%
which establishes an Luther-Emery liquid.


\end{document}